\documentclass[twocolumn]{aastex63}
\usepackage{graphics,epsf}
\usepackage{amsmath}                
\usepackage{amsfonts}               
\usepackage{amssymb}                
\usepackage{epsfig}                 
\usepackage{appendix}
\usepackage{graphicx}
\usepackage{float}
\usepackage{color}
\usepackage{multirow}
\usepackage{colortbl}
\usepackage[para,online,flushleft]{threeparttable}

\newcommand{\km}{{~\rm km}}
\newcommand{\s}{{~\rm s}}

\newcommand{\pc}{{~\rm pc}}

\newcommand{\AU}{{~\rm AU}}

\begin{document}

\title{A twin-jet structure rather than jet-rotation in the young stellar object OMC 2/FIR 6b}

\email{soker@physics.technion.ac.il}

\author{Noam Soker}
\affiliation{Department of Physics, Technion – Israel Institute of Technology, Haifa 3200003, Israel}

\author{Jesse Bublitz}
\affiliation{NRAO Green Bank Observatory, 155 Observatory Road, Green Bank, WV 24944, USA}

\author{Joel H. Kastner}
\affiliation{Chester F. Carlson Center for Imaging Science and Laboratory for Multiwavelength Astrophysics, Rochester Institute of Technology, 54 Lomb Memorial Drive, Rochester, NY 14623, USA}

\begin{abstract}
We analyse recent high-quality  Atacama Large Millimeter Array (ALMA) molecular line mapping observations of the northeast jet of the young stellar object (YSO) OMC~2/FIR~6b (HOPS-60) and find that these ALMA observations are much more likely to indicate a twin-jet structure than jet rotation,  as previously hypothesized. 
The interpretation of the line of sight velocity gradient across (perpendicular to its axis) the northeast jet of Fir~6b in terms of jet rotation leads to jet-launching radii of $\simeq 2-3 \AU$. However, the velocities of the jets $\simeq 100-400 \km \s^{-1}$ are much larger than the escape speed from these radii. 
We argue that the northeast jet of FIR~6b is instead compatible with a twin-jet structure, as observed in some planetary nebulae.
 Specifically, we find that the main, redshifted jet emanating from the central YSO is composed of two, very closely aligned, narrower jets that were launched by the central YSO at about the same time but at different inclinations with respect to the plain of the sky.
This twin-jet structure removes the extreme requirement that jets with velocities similar to the escape velocity from the YSO be launched from very large radii.
The YSO FIR~6b and  certain planetary nebulae also share  the characteristics of unequal structures and intensities of their two opposing bipolar jets. 
 We propose that such opposing lobe asymmetries can result from a sub-stellar binary companion on an eccentric orbit that is inclined to the accretion disk plane.  
\end{abstract}


\section{Introduction} 
\label{sec:intro}
    
There is a two-decade old dispute as to whether asymmetries in the line of sight velocities across some jets of young stellar objects (YSOs) are due to large scale rotations of the jets (e.g., \citealt{Bacciotti2002, Coffey2004, Coffey2007,  Lee2007, Lee2009, Chrysostomou2008, Zapata2010, Choi2011a, Pechetal2012, Chenetal2016, Hirotaetal2017}), or whether these asymmetrical perpendicular velocities are due to other effects. Such effects include interaction with the ambient gas (e.g., \citealt{Soker2010}), interaction with a twisted-tilted (wrapped) accretion disk (e.g., \citealt{Soker2005, Soker2007}), jet precession \citep{Cerqueira2006}, and twin-jet substructure \citep{SokerMcley2012}. In a different model, \cite{Fendt2011} proposes that magnetohydrodynamic shocks in a helical magnetic field transfer angular momentum to a jet that is launched with zero angular momentum. In general, the measurements of asymmetrical, perpendicular velocities are quite challenging and complicated (e.g., \citealt{Coffey2012}), and many effects might be misinterpreted as rotation (e.g., \citealt{Staffetal2015}).

The question of whether or not these asymmetrical perpendicular velocities are due to jet rotation has deep implications for the launching mechanisms of jets in YSOs and many other astrophysical objects, such as the progenitors of planetary nebulae (PNe). The studies that implicate jet rotation base their claims on the magneto-centrifugal acceleration model (e.g., \citealt{Anderson2003, Ferreira2006}), according to which the accretion disk launches the jets from an extended zone of radius $\approx 1 \AU - 10 \AU$ (a disk-wind model, e.g., \citealt{Pudritz2007}).  The other camp (e.g., \citealt{Soker2010}), on the other hand, argues that the launching takes place much closer to the central star, such that jet rotations are much slower and very difficult to measure. 

\cite{SokerMcley2013} analysed the kinematic observations of the YSO NGC~1333~IRAS~4A2 (\citealt{Choi2011a}) and compared this object to the pre-PN CRL~618, which has a set of `twin jets' emanating from each side of the central star (e.g., \citealt{Leeetal2003, Balicketal2013, Huangetal2016}). `Twin jets' are structures where the main jet actually consists of two very close and narrow jets that were launched at about the same time. \cite{SokerMcley2013} suggested that each of the two opposing jets of IRAS~4A2 is also composed of twin jets. 
They further proposed that, in both this YSO and CRL~618, a binary companion on an eccentric orbit causes the jet launching at the companion's periastron passages. 
They thus predict the presence of a low-mass stellar companion in both objects. In the case of the pre-PN CRL618, it is the companion that accretes mass and launches the jets, while in YSOs the companion perturbs the accretion disk around the primary star that launches the jets. We note that \cite{Velazquezetal2014} propose an asymmetrical jet-ejection mechanism to account for the jets of CRL~618.
  Some systems appear to have more than two pairs of opposite jets; e.g., \cite{Nakashima2010} and \cite{Derlopaetal2021}, respectively, have presented evidence that the planetary nebulae NGC~7027 and NGC~2818 each harbor three pairs of jets. 

Comparison of PNe and YSOs can inform us about both classes of objects (e.g., \citealt{LeeSahai2004}). Follow-up studies of PNe and pre-PNe have shown that some jets might have multi-jet structures that result from the periastron passage of a companion to the PN progenitor (e.g., \citealt{Sahai2016, Huangetal2020}). 
In addition, \cite{Kwonetal2015} present observations of the jets of the YSO  L1157 and claim that the outflow is consistent with multiple jets. 

In the present study we examine a recent claim for jet rotation in the YSO FIR~6b (HOPS-60), a protostar in Orion Molecular Cloud 2 \citep{Matsushitaetal2021}. We argue that it is much more likely that the asymmetrical perpendicular velocities of the red-shifted jet in this system reveals a twin-jet structure, rather than jet rotation (sections \ref{sec:Strucutre} and \ref{sec:OutflowVelocity}). This has implications for the jet-launching mechanism that we discuss in section \ref{sec:Mechanism}. 

\section{The structure of the red-shifted jet} 
\label{sec:Strucutre}

\cite{Matsushitaetal2021} conducted high-quality   Atacama Large Millimeter/submillimeter Array (ALMA) CO ($J = 2\rightarrow$1) and 1.3 mm continuum  observations of the YSO FIR~6b and present a detailed investigation of these data.
They included an analysis of the asymmetrical velocities perpendicular to the jet axis, and interpret these results in the frame of the magneto-centrifugal acceleration model of \cite{Anderson2003}. \cite{Matsushitaetal2021} discuss possible explanations other than jet rotation for the asymmetrical perpendicular velocities of the northeast (red-shifted) jet of FIR~6b, but dismiss them, concluding that ``...it is unlikely that the jet is composed of twin flows.'' 

\begin{figure*}
\begin{center}
\includegraphics[trim=9cm 12.6cm 9cm 2.00cm,scale=1.1]{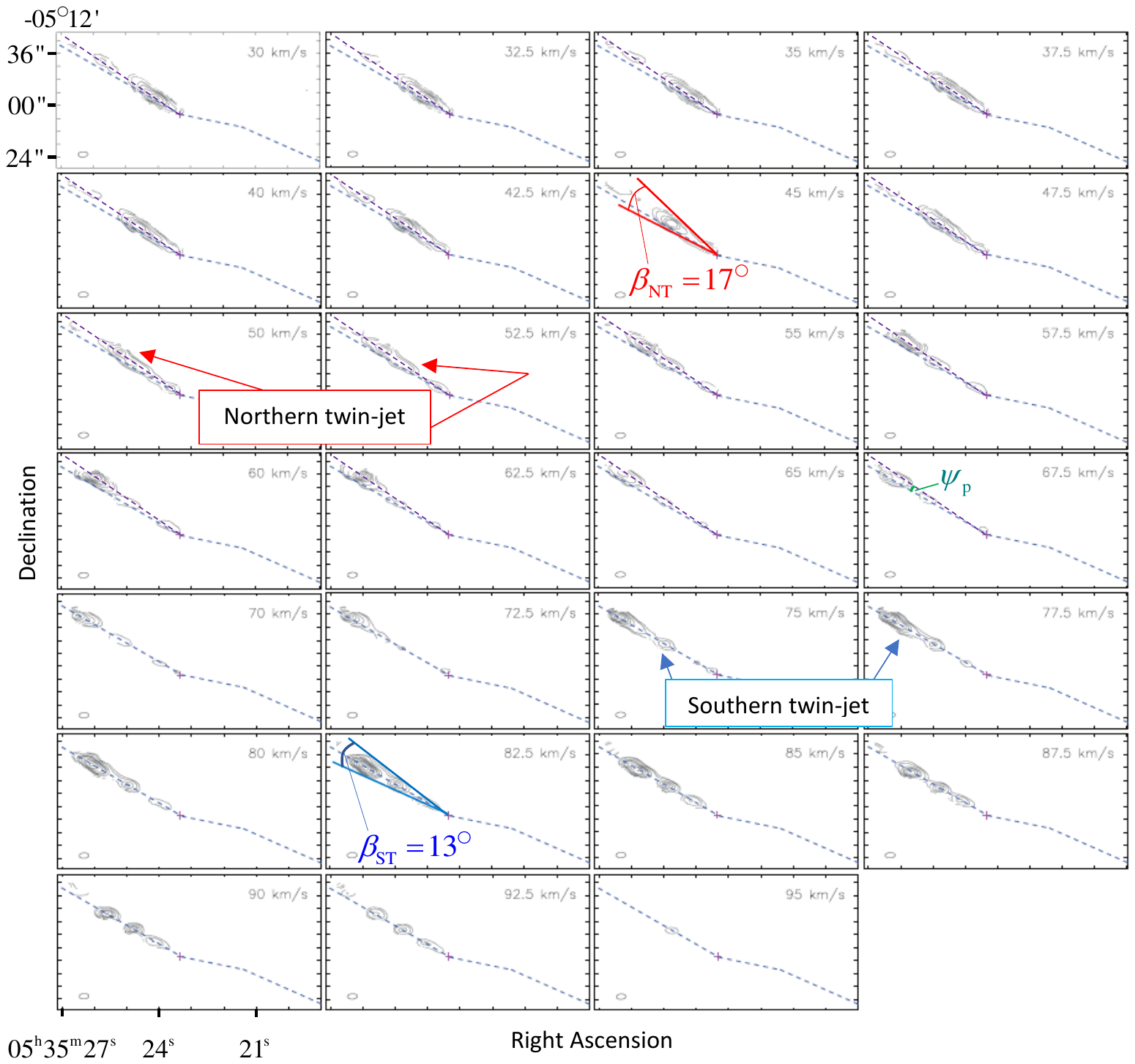}
\caption{  ALMA velocity (channel) map figure from \cite{Matsushitaetal2021} detailing CO ($J = 2\rightarrow$1) emission from the northeast outflow (jet) side of the bipolar outflow (the red-shifted jet) of FIR~6b, annotated to highlight features that are relevant to the present study. The dashed blue line that extends from upper left to lower right is from \cite{Matsushitaetal2021}, as is the red plus symbol denoting the position of the central star.  We have added a dashed purple line to mark the emission feature that we claim is the northern component of the red-shifted twin jet system, to highlight its displacement from what we claim is the southern component, which \cite{Matsushitaetal2021} marked with their dashed blue line. We also added annotations that indicate the opening angles of each of the two components of the twin jet ($\beta_{\rm NT}$ and $\beta_{\rm ST}$ for the northern and southern components of the twin jet, respectively), and the angle between the two components of the twin jet ($\Psi_P$), as projected on the plane of the sky. The velocities are with respect to the Local Standard of Rest (LSR) frame; the presented range of $30-95 \km \s^{-1}$ corresponds to a range of $19-84 \km \s^{-1}$ relative to the LSR velocity of the central YSO. For scaling, the distance to FIR~6b is $d \approx 393 \pc$ \citep{Tobinetal2020}. 
}
\label{fig:6panels}
\end{center}
\end{figure*}

\begin{figure*}
\begin{center}
\includegraphics[trim=0.0cm 0.0cm 0.0cm 0.0cm,scale=0.6]{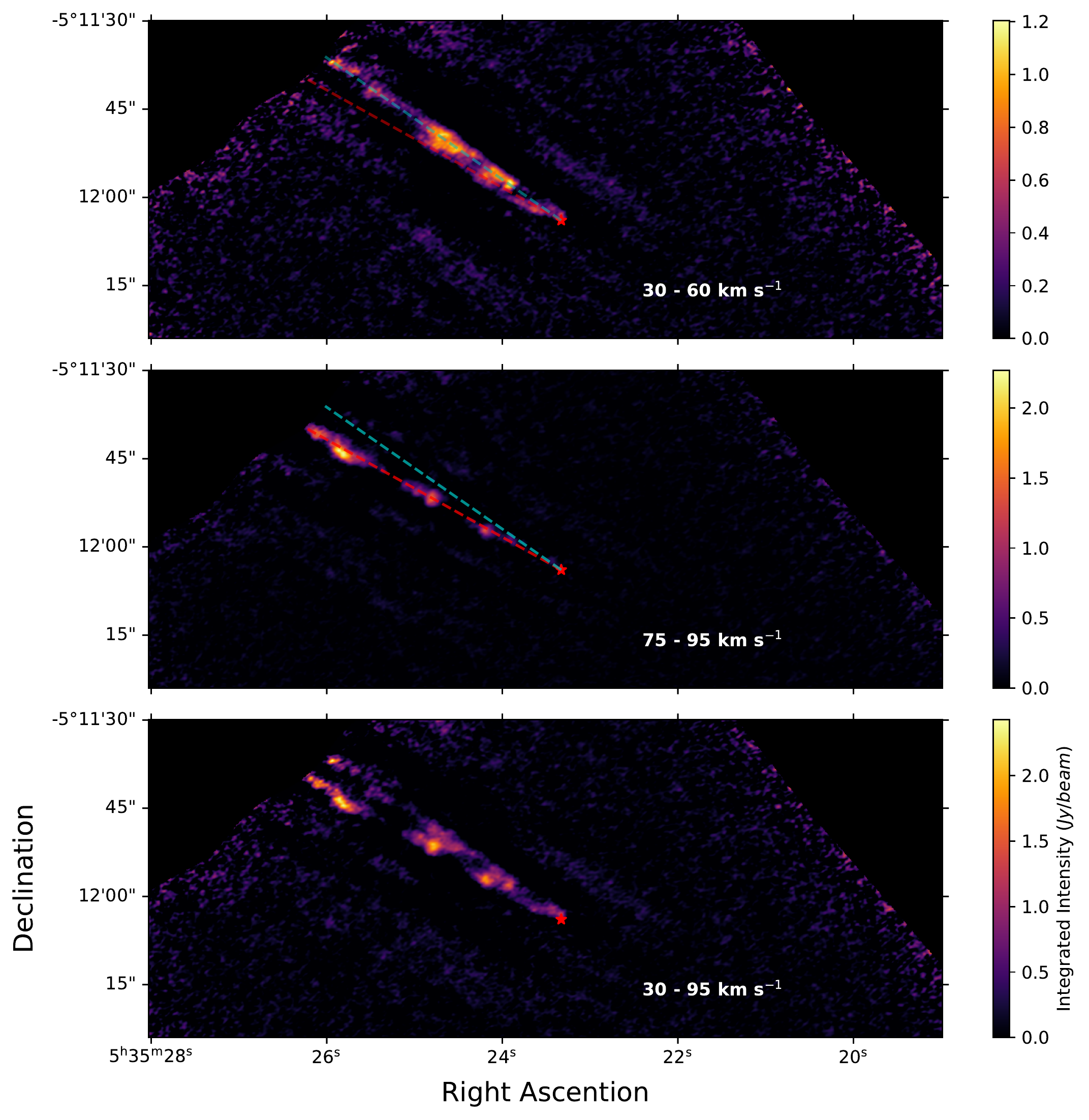}
\caption{   Our reconstruction of the ALMA CO emission maps from FIR~6b integrated over specific velocity ranges that isolate emission from the two (twin) components of the red-shifted jet. 
Top: map spanning the 30--60 km s$^{-1}$ velocity range, which isolates the northern twin-jet component. 
Middle: as in the top panel, for the 75--95 km s$^{-1}$ velocity range that isolates the southern twin-jet component. 
Bottom: map of the red-shifted jet integrated over the full velocity range spanned by the top and middle panels (30--95 km s$^{-1}$). In each panel, a red star marks the YSO, the putative source of the outflow, with the integrated velocity ranges indicated in the lower right. The blue and red dashed lines in the top and middle panels mark the trajectories of the northern and southern components of the twin-jet system, respectively. 
  }
\label{fig:NewVmap}
\end{center}
\end{figure*}

We here present evidence in support of this alternative view. 
  We first present our interpretation of their results (Fig. \ref{fig:6panels}), and then present new velocity maps of the red-shifted (northeast) jet (Fig. \ref{fig:NewVmap}).  

  In Fig. \ref{fig:6panels} we present our edited version of Figure 4 from \cite{Matsushitaetal2021}. We have added a straight dashed purple line to several panels to indicate our identification of the northern component of the twin jets that, together, comprise the red-shifted component of the bipolar outflow  extending northeast of the driving YSO. In the 45 km s$^{-1}$ (82.5 km s$^{-1}$) velocity panel we mark the opening angle of the northern (southern) components of the twin jet as projected on the plain of the sky and, in the 67.5 km s$^{-1}$ velocity panel, the projected angle between what we identify as the two components of the twin jet.  
Hence, in Fig. \ref{fig:6panels}, we can already clearly identify the northern and southern components of the twin jet as spanning different velocity ranges.

  In Fig. \ref{fig:NewVmap}, we present velocity-integrated maps obtained from our reanalysis of the ALMA CO data for FIR~6b presented in \cite{Matsushitaetal2021}, integrated over specific velocity ranges that isolate emission from the two components of the red-shifted twin jet system. 
Data sets for the CO ($J = 2\rightarrow$1) line observations were obtained from the ALMA archive, where they were processed via the ALMA pipeline, imaged with CASA, and continuum-subtracted. For the purposes of this analysis, the data were sufficiently clean to require no further processing.   
  
  It is readily evident from Fig. \ref{fig:NewVmap} (top and middle panels) that the northern component of the red-shifted twin jet is restricted to line of sight velocities in the range $ 30 \la v_{\rm NT,l} \la 64 \km \s^{-1}$, while its southern twin is restricted to the range $75 \km \s^{-1} \la v_{\rm ST,l} \la 95 \km \s^{-1}$. 

One argument that \cite{Matsushitaetal2021} present against a twin-jet structure is that there are no signatures of the twin jets in the integrated intensity map (their Figure 2). However,   in their Figure 2,  the ridge of highest intensity along the northeast outflow (jet) is clearly displaced to the south. We attribute this to the southern component of the twin jet system being brighter than the northern component. In addition,   in their Figure 2,  the southern component of the twin jet displays peaks at larger distances from the central YSO than the northern component, explaining why the ridge of the brightest emission shows increasing displacement to the south of the red-shifted outflow at increasing distances from the central YSO. 

From Figs. \ref{fig:6panels} and \ref{fig:NewVmap}, we estimate the projected angle between the two components of the twin-jet system to be 
$\Psi_{\rm p} \simeq 5^\circ$.  Based on Fig. \ref{fig:6panels}  we estimate the full opening angle of the northern component of the twin-jet system as $\beta_{\rm NT} \simeq 17^\circ$ and that of the southern component as $\beta_{\rm ST} \simeq 13^\circ$. Clearly, because $\psi_{\rm p} \ll 0.5(\beta_{\rm NT}+\beta_{\rm ST})$, the projections of the two components largely overlap with each other on the plane of the sky. This, and the fact that the southern twin-jet is much brighter, explains why it is not simple to identify the northern component of the twin-jet system in the integrated intensity map of \cite{Matsushitaetal2021}. 
  Indeed, the map in the bottom panel of our Fig. \ref{fig:NewVmap}, which is integrated over the full red-shifted emission velocity range (30--95 $\km \s^{-1}$), clearly reveals both components of the twin-jet system. 

To estimate the three-dimensional (3D) angle between the two components of the twin jet, we would need the inclination angle of the red-shifted outflow with respect to the line of sight. However, the inclination angle of the disk around the star is not well determined; \cite{Matsushitaetal2021} cite a wide range of possible inclinations, from $i \simeq 42^\circ$ to $i \simeq 80^\circ$. Based on these different values, \cite{Matsushitaetal2021} build three models in which the velocities of the jets range from $v_{\rm j} \simeq 80 \km \s^{-1}$ in one of the three models to $v_{\rm j} \simeq 410 \km \s^{-1}$ in their preferred model. 
We assume here  $v_{\rm j} = 200 \km \s^{-1}$, for purposes of demonstrating the possible 3D twin-jet structure. 

From the velocity maps of \cite{Matsushitaetal2021}, we find the peak intensity of the northern component of the twin-jet system arises at a line of sight velocity of $v_{\rm NT,l} \simeq 60 \km \s^{-1}$. 
This is also the emission peak that lies at the largest distance from the central YSO. For the southern component, the line of sight velocity of peak emission is $v_{\rm ST,l} \simeq 82.5 \km \s^{-1}$. These velocities are 
$v^\ast_{\rm NT,l} \simeq 49 \km \s^{-1}$ and $v^\ast_{\rm ST,l} \simeq 71.5 \km \s^{-1}$, respectively, with respect to the systemic (central YSO) velocity.
For a jet velocity of $v_{\rm j} = 200 \km \s^{-1}$, the inclination angles of the two components of the twin jet with respect to the line of sight are hence $i_{\rm NT} \simeq 76^\circ$ and
$i_{\rm ST} \simeq 69^\circ$. Overall,  considering the uncertainties, including the $\simeq 1 ^\circ$ uncertainty in $\psi_{\rm p}$,  we estimate the true angle between the axes of the two jets to be $\psi \simeq 8^\circ - 9^\circ$.  This is smaller than the sum of the half-opening angles of the two jets, i.e., $\psi < 0.5(\beta_{\rm NT}+\beta_{\rm ST}) \simeq 15^\circ$, as deduced from the velocity maps of  \cite{Matsushitaetal2021} (Figure~\ref{fig:6panels}). This conclusion comes from our assumption that the two jets share the same velocity of $v_{\rm j}=200 \km \s^{-1}$. However, most likely their velocities will be different, which might affect the interpretation of their precise spatial relationship.   If indeed the jets extend to these widths, the inequality implies that the two jets may interact with each other at their outskirts.  Nonetheless, our new velocity maps clearly reveal the presence of a twin-jet system, wherein the red-shifted jet to the NW of the driving YSO has distinct northern and southern components, and suggests that the interaction between these two components does not significantly influence their expansion after launching. 

\section{Jet energetics} 
\label{sec:OutflowVelocity}

\cite{Matsushitaetal2021} determine the velocity gradient at nine positions across the northeast jet (asymmetrical perpendicular velocities)  via cuts across the jet. They then apply the model of \cite{Anderson2003} to determine the launching radius of each of these nine positions along the jet. In the outer four positions that they analyse, the jet is wide and has a large surface brightness, so the noise is low. For these four positions the outflow jet velocities (which they term $v_{\rm pol}$) are $v_{\rm j} \simeq 402-410 \km s^{-1}$ and the launching radii are $r_0 \simeq 2.16-2.60 \AU$. For their assumed stellar mass of $M_{\rm star} = 1 M_\odot$, the escape velocities from these radii are $v_{\rm esc} (r_0) \simeq 29-26 \km \s^{-1}$. Hence,   according to their analysis,  the jet velocity would be roughly 14 times larger than the escape velocity from their estimated launching radii. 

\cite{Matsushitaetal2021} refer to the large angular momentum of the jet   under the rotating jet scenario and note that such a large angular momentum requires  a very large lever-arm, $\approx 14 r_0$, in the framework of the \cite{Anderson2003} model. 
We raise here the question of energy, or escape velocity. In all classes of astrophysical jets that are launched by mass-accreting objects, observations indicate that the typical jet velocities are on the order of the escape velocity from the mass-accreting object (e.g., \citealt{Livio1999}). This suggests that the typical jet-launching radius is on the order of the radius of the central object. 
Indeed, in the case of FIR~6b the jet velocities in the three models that \cite{Matsushitaetal2021} consider are roughly that of the escape velocity from a low-mass YSO. We take this to imply that the launching radius is $\approx 1-5 R_\odot$;    whereas, in the disk wind model adopted by \cite{Matsushitaetal2021}, the launching radius would be a hundred times larger, i.e., $\approx 1-10$ au.  

  Given the above, we can estimate the parameters of the jets that \cite{Matsushitaetal2021} are invoking using the detailed modelling of magneto-thermal disk winds from protoplanetary disks by \cite{Baietal2016}. 
This model includes the dimensionless ``mass loading'' parameter $\mu$ (their equation 18) that determines the ratio of the mass loss rate into the wind, $\dot M_{\rm wind}$, relative to the accretion rate, $\dot M_{\rm acc}$. For the lever arm value of $\approx 14 r_0$ that \cite{Matsushitaetal2021} require, the range of $\mu$ 
is $\mu \approx 7 \times 10^{-4} - 3 \times 10^{-3}$, given the parameter space explored by \cite{Baietal2016}. For a wind velocity at infinity that is about 20 times the Keplerian velocity at the foot-point of the wind, as required by the rotating jet in the \cite{Matsushitaetal2021} scenario, the value of $\mu$ is in the range $\mu= 10^{-4} - 6 \times 10^{-4}$. The mass loss rate into the wind for this range of parameters is  $\dot M_{\rm wind}/\dot M_{\rm acc} \simeq (\mu /4) < 0.001$. In contrast, observations suggest $(\dot M_{\rm wind} /\dot M_{\rm acc} )_{\rm observed} \approx 0.1-1$, with large uncertainties \citep{Baietal2016}. 
We conclude that although the dynamical properties of the jets that \cite{Matsushitaetal2021} propose are possible, the mass loss rate into the jets would have to be two orders of magnitude below typically observed values.

As shown above, the model invoked by \cite{Matsushitaetal2021} also results in jet velocities that are more than an order of magnitude larger than the escape velocity from the launching radii. 
Furthermore, one would expect that the extremely strong magnetic field required to maintain very large lever arms would also result in two opposite bipolar outflows that are similar in shape and intensity to each other; whereas, as demonstrated by \cite{Matsushitaetal2021}, the two flows in FIR~6b are highly asymmetric. 

To summarise this short section, in the magneto-centrifugal acceleration model there is no observational basis to introduce a scale of several au --- i.e., two orders of magnitude larger than the stellar radius --- for the jet launching radii, nor any mechanism that might explain the largely unequal sides of the bipolar outflow.  
 We claim that another process must take place in this system.
 
\section{Implications for the jet-launching mechanism} 
\label{sec:Mechanism}

We have re-investigated the high-quality observations by \cite{Matsushitaetal2021} of the YSO FIR~6b (HOPS-60) and analysed their claim that the asymmetrical perpendicular (to the jet axis) line of sight velocities of the northeast jet emanating from this YSO are due to the rotation of the jet around its axis. Instead, we attribute the asymmetrical perpendicular velocities to a twin-jet structure of this outflow (Figs.  \ref{fig:6panels} and \ref{fig:NewVmap}) rather than to jet rotation. 

The highly unequal red-shifted (northeast) and blue-shifted (southwest) bipolar outflows of FIR~6b are qualitatively similar to the highly unequal bipolar lobes of the pre-PN CRL 618 (e.g., \citealt{Leeetal2003, Balicketal2013, Huangetal2016}).  { The influence of a binary companion is thought to generate such bipolar structure in PNe and pre-PNe (e.g., \citealt{DeMarco2009}). CRL 618's lobes clearly display the structure of twin jets on each side of the central, evolved (post-asymptotic giant branch) star. } This asymmetrical structure of unequal lobes suggests that the accretion disk that has launched the jets was strongly disturbed during the launching process. A binary companion on an eccentric orbit that is inclined to the accretion disk plane might generate such a non-axisymmetrical perturbation \citep{SokerMcley2013}. 

 { By analogy with the  \cite{SokerMcley2013} model for the asymmetrical twin-jet structure of CRL~618,} we here suggest that a substellar object (a brown dwarf or a massive planet) on an eccentric orbit inclined to the accretion disk plane orbits the YSO FIR~6b. This substellar object is responsible for the twin-jet structure of the northeast outflow and the inequality of the two opposite sides of the bipolar outflow, via its perturbing interactions with the jet-launching accretion disk of the YSO. 

The present short study has significant implications for the jet launching process: the twin-jet structure that we presented here removes the extreme requirement that jets with velocities similar to the escape velocity from the YSO be launched from very large radii of $r_0 \ga 2 \AU$, where the escape velocity is more than an order of magnitude smaller than the jet velocity. Instead, in the model proposed here, the YSO launches the jets from its vicinity, $\simeq 1-5 R_\odot$, where the escape velocity is very similar to the observed velocities of the jets.
We also note that a twin-jet structure where two jets are launched separately can more easily account for the unequal intensities of the two jets than can a model invoking the rotation of a single jet.
Repeat ALMA observations in about a decade might allow us to determine the difference in proper motion between the northern and southern components of the red-shifted twin jet, from which we might refine the twin-jet model proposed here.

\section*{Acknowledgments}

We thank the referee, Pat Hartigan, for very useful comments and suggestions that led to significant improvements in the manuscript. NS was supported by a grant from the Israel Science Foundation (769/20) and by the Pazy Research Foundation. 



\label{lastpage}
\end{document}